# Noncoding RNAs serve as the deadliest regulators for cancer

Anyou Wang[1*], Hai Rong[1,2]

[1]The Institute for Integrative Genome Biology, University of California at Riverside, Riverside, CA 92521, USA, and [2]Department of Microbiology and Plant Pathology, University of California at Riverside, Riverside, CA 92521, USA

[*]Correspondence:
A Wang anyou.wang@alumni.ucr.edu

**Abstract:** Cancer is one of the leading causes of human death. Many efforts have made to understand its mechanism and have further identified many proteins and DNA sequence variations as suspected targets for therapy. However, drugs targeting these targets have low success rates, suggesting the basic mechanism still remains unclear. Here, we develop a computational software combining Cox proportional-hazards model and stability-selection to unearth an overlooked, yet the most important cancer drivers hidden in massive data from The Cancer Genome Atlas (TCGA), including 11,574 RNAseq samples and clinic data. Generally, noncoding RNAs primarily regulate cancer deaths and work as the deadliest cancer inducers and repressors, in contrast to proteins as conventionally thought. Especially, processed-pseudogenes serve as the primary cancer inducers, while lincRNA and antisense RNAs dominate the repressors. Strikingly, noncoding RNAs serves as the universal strongest regulators for all cancer types although personal clinic variables such as alcohol and smoking significantly alter cancer genome. Furthermore, noncoding RNAs also work as central hubs in cancer regulatory network and as biomarkers to discriminate cancer types. Therefore, noncoding RNAs overall serve as the deadliest cancer regulators, which refreshes the basic concept of cancer mechanism and builds a novel basis for cancer research and therapy. Biological functions of pseudogenes have rarely been recognized. Here we reveal them as the most important cancer drivers for all cancer types from big data, breaking a wall to explore their biological potentials.

## Introduction

Cancer has been a leading cause of death globally. In 2018, more than 609K patients died in US. By 2030, annual new cancer cases will rise to 23.6M worldwide[1]. It is vital and urgent to understand its mechanisms toward therapy.

The current researches on cancer mechanism have mostly focused on protein-coding regions, both functional genomics and genome sequences, and have revealed hundred proteins as key oncogenes and have identified thousands of sequence variations across different cancer types[2–6]. Many alterations in chromosomes and phenotypes have been observed[7]. Recently studies also have explored cancer epigenetics mechanisms[8] and have identified a certain groups of noncoding RNAs associated with cancer development[9] . These mechanisms have helped to improve technologies on therapy[10]. However, cancer is still an incurable disease and drugs targeting molecular targets based on current mechanisms are rarely successful[11,12], suggesting that the real fundamental mechanism still remains unclear.

Here, we developed a computational software to unearth the basic cancer mechanism hidden in the massive data from TCGA database.

## Results

### Noncoding RNAs serve as the deadliest regulators of cancers

To investigate the deadliest regulators for cancers, we needed a software that could unbiasedly reveal the strongest genes corresponding to cancer death and a data set including all types of cancers. Conventionally, the Cox proportional-hazards regression model (coxph) has been widely employed to find cancer genes[13], but it lacks random sampling during regression, leading to inaccurate estimations. Here, we developed an improved survival analysis software, referred as ISURVIVAL, to insert stability-selection[14] into coxph model to make results accurate (materials and methods). Implementing ISURVIVAL with julia computer language also makes it much faster than current software. The data used in this study was downloaded from The Cancer Genome Atlas (TCGA) public database, which includes 11,574 RNAseq samples and clinic data for all 36 cancer types (**Table S1**), which measured total 60,483 genes annotated with GRCh38.p2.v22 by TCGA (materials and methods).

To select the deadliest genes, we applied ISURVIVAL to a data matrix containing clinic time and status and each gene RNAseq data in all cancer samples and selected the top 428 significant genes with absolute coefficient >1 and pvalue<1.e-10 (materials and methods). We further ranked these 428 regulators in basis of their absolute coefficients (coef), higher coefficient, more important in regulating cancer death. Among 428, 92% were cancer inducers (coef > 1, HR>2.72), and only less than 8% were cancer repressors (coef< -1, HR< 0.37) **(Figure 1A)**. Evermore, all top 30 regulators were 100% cancer inducers. Interestingly, processed-pseudogenes (p-pseudogenes) occupied more than 80% in these top 394 inducers (**Figure 1B** left panel). At top 10 inducers, 6 (60%) were p-seudogenes, and top 1 was PANDAR (lincRNA) (**Figure 1B right**). Consistently, previous studies also reported PANDAR as a caner inducer[15]. Among cancer repressors, lincRNAs and antisense RNA occupied 40% and 30% respectively**(Figure 1C)**. This suggested noncoding RNAs as the deadliest cancer regulators and pseudo-genes as primary inducers and lincRNA and antisense RNA as key repressors. This parallels with our recent study on cancer mechanism revealing noncoding RNAs as the core drivers for all types of cancer[16].

**Personal clinic variables account for variations of activated regulators**

Researchers on cancer DNA sequences have attempted to find the genes with consensus variations across cancer types (disease types as described in table S1), but could not find any genes with more than 50% common mutations[6]. To test the consistence of gene activation across cancer types, we computed the coefficient of each gene in each cancer type, and found no consistence but high variation in the regulator activation. Even the top 10 inducers and repressors identified above Figure 1, only <60% and <67% of inducers and repressors were respectively activated as inducers (coef>0) and repressors (coef<0) across cancer types (**Figure 2**A). For example, the top 1 inducer, PANDAR, did not always work as an inducer, but as an inducer in only 57% cancer types and as a repressor in the rest.

We tried to find the reason for the activation variation, and examined the clinic variable effect on the variations of top 30 regulators by using canonical correspondence analysis (CCA). Clinic variables significantly affect the activation of genes. Among 7 character variables, sample type and alcohol affected most (**Figure 2B**), while smoking and BMI as the digital variables accounted for most of variances (**Figure 2C**). For example, PANDAR was positively correlated with sample type and negative correlated to race **(Figure 2B)**, significantly higher in primary tumor (p=5.8e-9, t-test, **Figure S1**) and Asian (p=0.00019, **figure S2**). However, PANDAR did not seemed to respond very well cancer types and tissue (**Figure 2B**). Yet tissue type was used to classify cancer types. This at least partially explain its activation variation (only 57%) in current classification system-disease type .

This observation above encouraged us to extend our CCA analysis to all genes and all 11 clinic variables. The whole cancer genome was clearly separated into three sections. The first positively affected by alcohol and smoking, the second by site, BMI and disease type, the third by unidentified factors **(Figure 2**D). Astonishingly, alcohol served as the strongest factor altering cancer genome activation, even stronger than smoking and BMI (**Figure 2D**). This indicated that the cancer genome activation is not only based on tissues as practiced by current classification system, but, importantly, based on other personal clinical variables. Therefore, using the current cancer type info to evaluate the activation of inducers and repressors as shown in Figure 2A could be misleading. Furthermore, even the first 200 principal components (PCs) derived from principal component analysis (PCA) only accounted for 60% variances (**Figure 2E**), indicating cancer genome activation varying with unexpectedly complicated personal clinic variables.

**Noncoding RNAs as universal drivers for cancers**

Cancer regulator activation varies with complex factors as observed above. To search the regulators universal for all types of cancers, we must concern personal clinic variables. Here, we inserted the all 11 clinic variables and the first 200 PCs as covariates into our ISURVIVAL to select regulators general for all cancers. The top 64 significant regulators ($p<0.05$, HR >1.1 or HR <0.9) included 65.6% inducers (HR>1.1) and 34.4% repressors (HR<0.9)(**Figure 3 A**). Among inducers, p-pseudogenes dominated all profile and occupied 40% of top 10 (**Figure 3B**). As expected, these inducers were consistently activated as inducers in most cancer types, with RP11-335k5.2 activated in 83% cancer types (**Figure 3C**). The overall percentage of inducer activation was significantly higher than previous one without concerning personal clinic variables shown in Figure 2A ($p = 0.03151$ and 0.08232 respectively for top 20 and top 10 inducers, t-test). This indicated these pseudogene-dominated inducers as universal inducers of cancers.

All universal repressors were also noncoding RNAs, including p-pseudogenes, lincRNAs, TEC, antisense and sense_intronic (**Figure S3**). Together suggested noncoding RNAs as the universal drivers for all cancers.

## Noncoding RNAs serve as the deadliest hubs in cancer regulatory network

We previously established a cancer regulatory network and computed the centrality as top hubs in cancer genome. Here, we filtered these centrality with survival data ($p<0.01$ and $HR<0.9$ or $HR>1.1$) and obtained the deadliest hubs. More than 75% of top 491 hubs were inducers (**Figure 4A**) and all top 50 inducers as p_pseudogenes (**Figure 4B**). Moreover, lincRNAs, antisense, p_ pseudogene and sense_intronic dominated the top 50 repressors (**Figure 4C**), suggesting noncoding RNAs as the deadliest centrality in cancer genome.

## Noncoding RNAs as cancer biomarkers

We further examined whether noncoding RNAs work as biomarkers to discriminate cancer and normal samples. We used total 632 normal samples as normal to discriminate cancer samples of each cancer type, and employed elastic-net with stability-selection to select biomarkers for each cancer type (materials and methods). Typically, around 50 noncoding RNAs could discriminate cancer vs normal with rare error (**Figure 5A**). We used the top 50 noncoding RNAs in each cancer type as biomarkers to calculate the discrimination accuracy by computing AUC (Area Under The Curve) of ROC (Receiver Operating Characteristics). Beginning with top 2 biomarkers, the AUC of each cancer type reached >90%, and with top 20 biomarkers, AUC reached stable state (>96%) for all cancer types (**Figure 5B**). This indicated noncoding RNAs can be used as biomarkers to discriminate cancer types.

## Discussion

Current researches on cancer drivers and mechanisms have focused on proteins and protein-relevant DNA sequences[2]. However, recent observations suggested most of these proteins as incorrect targets, which were selected from current mechanism studies [11]. This suggests that the current mechanisms are misleading. Here, we first time unearthed noncoding RNAs as the universal deadliest drivers for all types of cancers. Among noncoding RNAs, p-pseudogenes work as the primary cancer inducers. Noncoding RNAs have been reported as regulators for cancer [17], and p-pseudogenes mutations have been found in cancers [18], but they have not been reported as the key players replacing the protein spot in cancers. Recent reports regarding noncoding RNAs have focused on lincRNA category but have rarely demonstrated these pseudo-gene roles in cancer. Our findings here have established pseudogenes as the most important cancer regulators, instead of proteins as conventionally thought. This provides a novel mechanism and targets for cancer research and therapy and eventually make cancer curable.

One of long standing puzzle hanging in cancer researches is the universality of oncogenes. Many efforts have put into identifying the sequence consensus among these oncogenes, but no oncogene has more than 50 consensus sequences across cancer types [6]. Here, we revealed that the universal oncogenes exist in noncoding RNAs after normalizing personal clinic variables. For example,

RP11-335k5.2 was consistently activated in 83% cancer types although current cancer type classification is misleading as described below.

Cancer types classified by tissue types have been used in cancer researched and therapy development[2], yet cancer genome activation does not respond very well to tissues as shown in the present study. Surprisingly, alcohol contributes to cancer genome variations more than smoking, and these personal clinic variables such as alcohol and smoking are more important than tissue types in altering cancer genome activation. Therefore, the current cancer classification system may be misleading and should be reviewed to include personal variables to make therapy efficiently.

The current personal medicine on cancer has focused on personal DNA sequence variation[19 20], but personal clinic variables such as alcohol, smoking and BMI significantly alter cancer genome activation. These personal variables may be more important than personal DNA sequence variation in personal medicine.

Overall, our finding based on personal clinic variables and gene expression data has established noncoding RNAs as the most important inducers and repressors for all cancers. This refreshes the concept on cancer mechanism and therapy.

## Figure legends

**Figure 1. The deadliest regulators in cancers**. A, the proportion (%) of strongest inducers and repressors. B, gene categories of top deadliest inducers (left panel) and top 10 inducers (right). For clear illustration, only top 5 gene categories were shown here and thereafter in this study. C, gene categories of top repressors (left) and top 10 repressors.

**Figure 2. Personal clinic variables account for variations of cancer gene expressions.** A, heatmap shows gene coefficent variations of top 10 inducers (upper) and repressors (bottom) across 30 individual cancer types. Each row denotes a gene and each column as a cancer type. For illustration propose, the coefficient of a gene in a cancer type was set to 1 (coefficient >0, red color) or -1 (coefficient <0, blue color) or 0 (coefficient=0, white). The number following gene symbol represents the frequency (%) of this gene in 30 cancer types as an inducer (upper) or a repressor (bottom). For example, 0.6 in ACTG1P11 = 18(red as inducer)/30. Left bar colors represent gene categories. In inducer panel(upper), blue: processed-pseudogenes (p_pseudogene, thereafter), red:lincRNA, green:protein_coding. In the bottom repressor, pink:TEC, red:antisense, blue:linRNA, green:p_pseudogene, brown: protein_coding (protein, thereafer). B, Canonical correspondence analysis (CCA) plot of top 30 regulators and 7 character clinic variables from all cancer samples. C, CCA plot of top 30 regulators and 4 digital clinic variables from all cancer samples. D, CCA plot of all genes and total 11 clinic variables from all cancer samples. Cancer genome was clearly seperated into 3 clusters based on clinic variables. E, The cumulative variances accounted

by each principal component (PC), from PC1 to PC200, derived from principal component analysis (PCA) of expressions of all genes and all cancer samples.

**Figure 3. Universal cancer inducers and repressors.** A, proportion of universal inducers and repressors. B, Gene categories of inducers. C, coefficient heatmap of top 10 universal inducers in 30 cancer types. Left bar colors denote gene categories, red:antisense, brown:p-pseudogene, blue:lincRNA, pink:protein, green:miRNA, yellow:snRNA.

**Figure 4. Distribution of inducers and repressors inferred from genome regulatory network**. A, proportion of top inducers and repressors derived from network centrality. B, Gene categories of inducers. C, gene categories of repressors.

**Figure 5. noncoding RNAs as biomarkers to discriminate 30 cancer types**. A, a typical cross-validation curve of noncoding RNAs as biomarkers. When total around 50 noncoding RNAs were selected, the mean cross-validated error reached minimum, and this corresponding lambda was selected as lambda.min for biomarker selections for all 30 cancer types. B, AUC of noncoding RNAs as biomarkers to discriminate 30 cancer types. When 10 noncoding RNAs as biomarkers, AUCs for most cancer types have not reached stable states, but when 20 noncoding RNAs were selected as biomarkers, all AUCs were stable, with AUC>0.96.

Fig.1

A

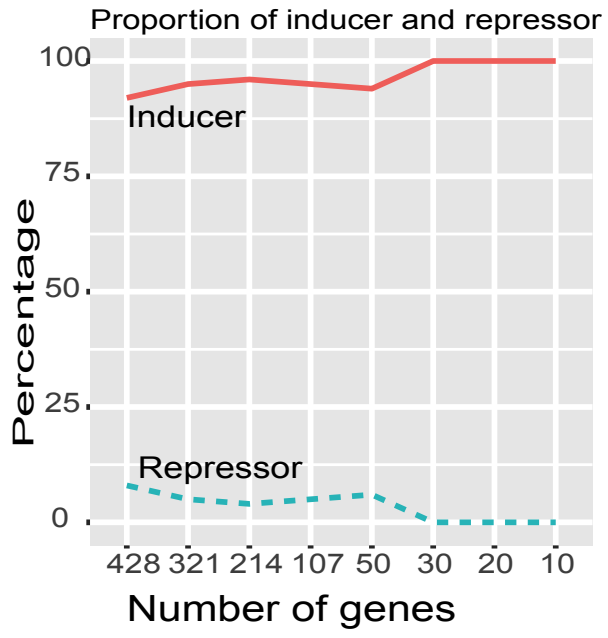

B

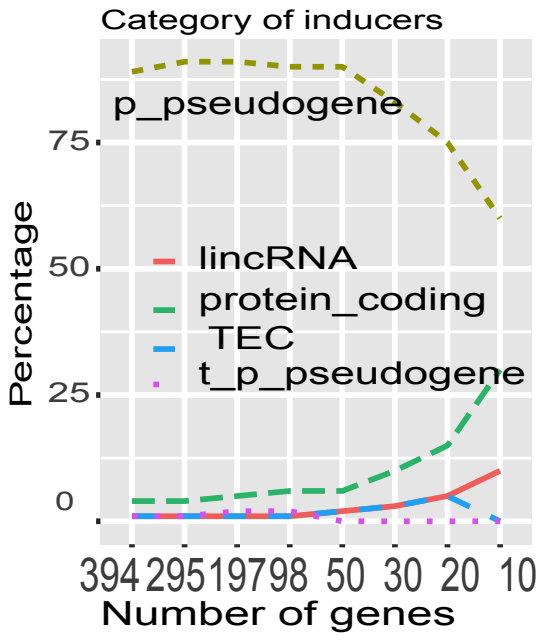

**Top 10 inducers**

| symbol | category | coef |
|---|---|---|
| PANDAR | lincRNA | 94.96 |
| ATF4P1 | processed_pseudogene | 89.67 |
| RGPD6 | protein_coding | 81.62 |
| PPIAL4F | protein_coding | 52.87 |
| ACTG1P2 | processed_pseudogene | 51.06 |
| BCLAF1P1 | processed_pseudogene | 38.90 |
| PARP1P1 | processed_pseudogene | 33.73 |
| ACTG1P11 | processed_pseudogene | 33.26 |
| RP11-340I6.6 | processed_pseudogene | 31.60 |
| PPIAL4D | protein_coding | 29.14 |

C

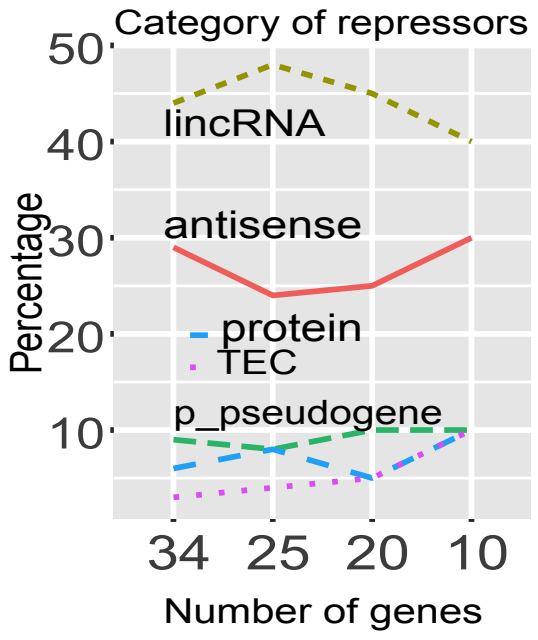

**Top 10 repressors**

| symbol | category | coef |
|---|---|---|
| RP11-109E24.2 | antisense | -15.93 |
| CTD-2515H24.2 | lincRNA | -15.16 |
| RP11-325I22.3 | TEC | -14.89 |
| LOC401913 | processed_pseudogene | -11.23 |
| RP11-98D18.16 | antisense | -9.70 |
| RAB6C | protein_coding | -5.96 |
| RP1-178F10.1 | antisense | -5.43 |
| RP11-285E23.2 | lincRNA | -5.19 |
| RP11-46H11.3 | lincRNA | -4.68 |
| SLC25A30-AS1 | lincRNA | -4.28 |

Fig2

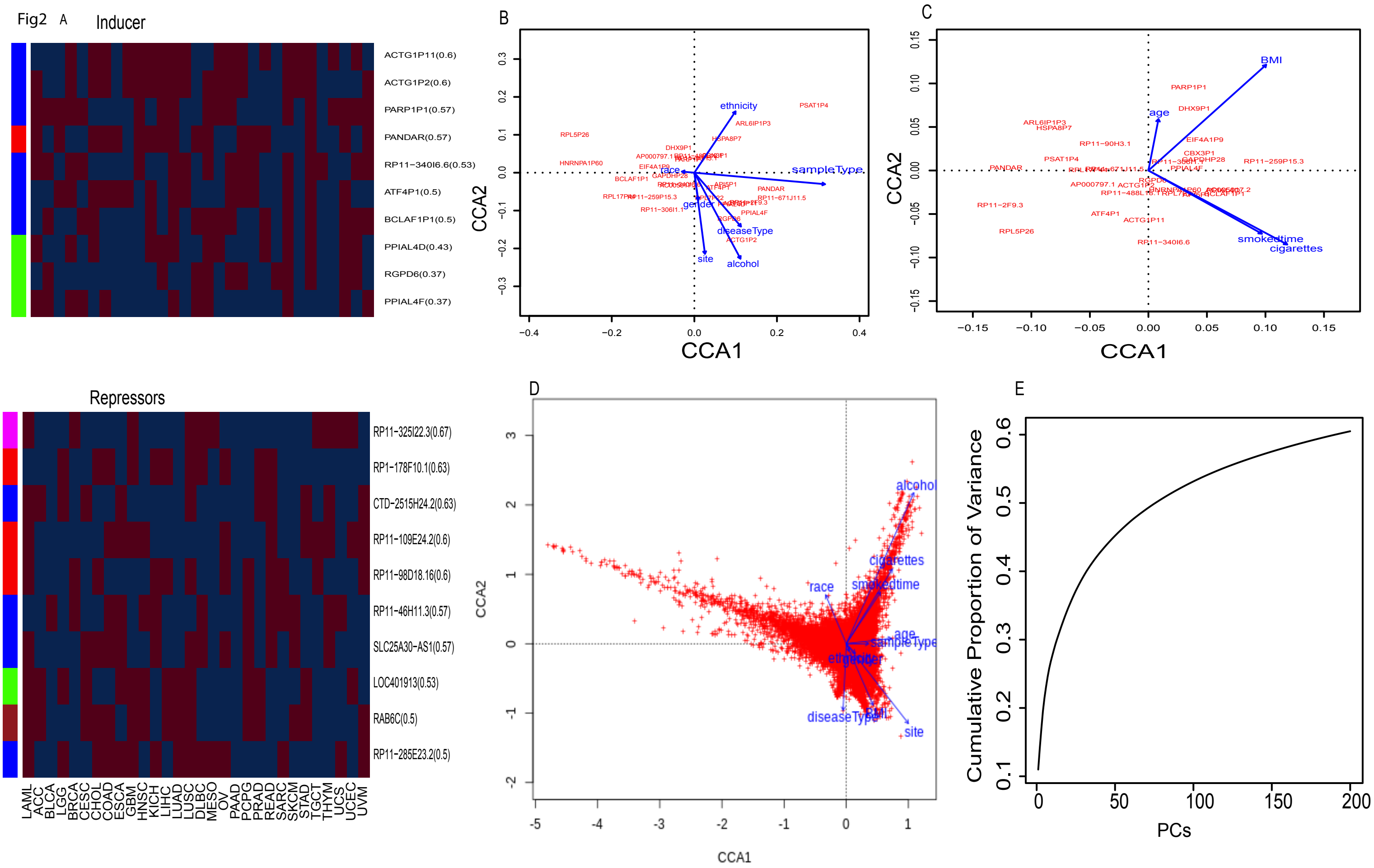
A
Inducer
ACTG1P11(0.6)
ACTG1P2(0.6)
PARP1P1(0.57)
PANDAR(0.57)
RP11−340I6.6(0.53)
ATF4P1(0.5)
BCLAF1P1(0.5)
PPIAL4D(0.43)
RGPD6(0.37)
PPIAL4F(0.37)
Repressors
RP11−325I22.3(0.67)
RP1−178F10.1(0.63)
CTD−2515H24.2(0.63)
RP11−109E24.2(0.6)
RP11−98D18.16(0.6)
RP11−46H11.3(0.57)
SLC25A30−AS1(0.57)
LOC401913(0.53)
RAB6C(0.5)
RP11−285E23.2(0.5)
LAML ACC BLCA LGG BRCA CESC CHOL COAD ESCA GBM HNSC KICH LIHC LUAD LUSC DLBC MESO OV PAAD PCPG PRAD READ SARC SKCM STAD TGCT THYM UCS UCEC UVM
B
CCA1
CCA2
ethnicity
sampleType
race
gender
diseaseType
site
alcohol
PSAT1P4
ARL6IP1P3
HSPA8P7
RPL5P26
HNRNPA1P60
DHX9P1
BCLAF1P1
PANDAR
RP11−671J11.5
PPIAL4F
ACTG1P2
RP11−259P15.3
RP11−308I1.1
C
CCA1
CCA2
BMI
age
smokedtime
cigarettes
PARP1P1
DHX9P1
ARL6IP1P3
HSPA8P7
EIF4A1P9
RP11−90H3.1
CBX3P1
PSAT1P4
RP11−259P15.3
PANDAR
AP000797.1
ATF4P1
ACTG1P11
RP11−2F9.3
RPL5P26
RP11−340I6.6
D
CCA1
CCA2
alcohol
cigarettes
race
smokedtime
age
sampleType
diseaseType
site
E
Cumulative Proportion of Variance
PCs

Fig3

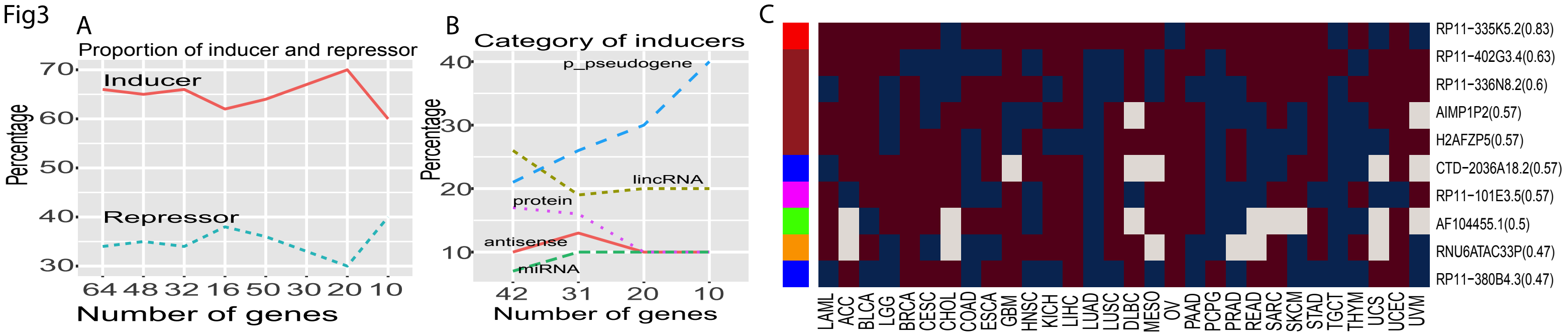
A
Proportion of inducer and repressor
Inducer
Repressor
Percentage
70
60
50
40
30
64 48 32 16 50 30 20 10
Number of genes
B
Category of inducers
p_pseudogene
lincRNA
protein
antisense
miRNA
Percentage
40
30
20
10
42 31 20 10
Number of genes
C
RP11-335K5.2(0.83)
RP11-402G3.4(0.63)
RP11-336N8.2(0.6)
AIMP1P2(0.57)
H2AFZP5(0.57)
CTD-2036A18.2(0.57)
RP11-101E3.5(0.57)
AF104455.1(0.5)
RNU6ATAC33P(0.47)
RP11-380B4.3(0.47)
LAML ACC BLCA LGG BRCA CESC CHOL COAD ESCA GBM HNSC KICH LIHC LUAD LUSC DLBC MESO OV PAAD PCPG PRAD READ SARC SKCM STAD TGCT THYM UCS UCEC UVM

Fig4

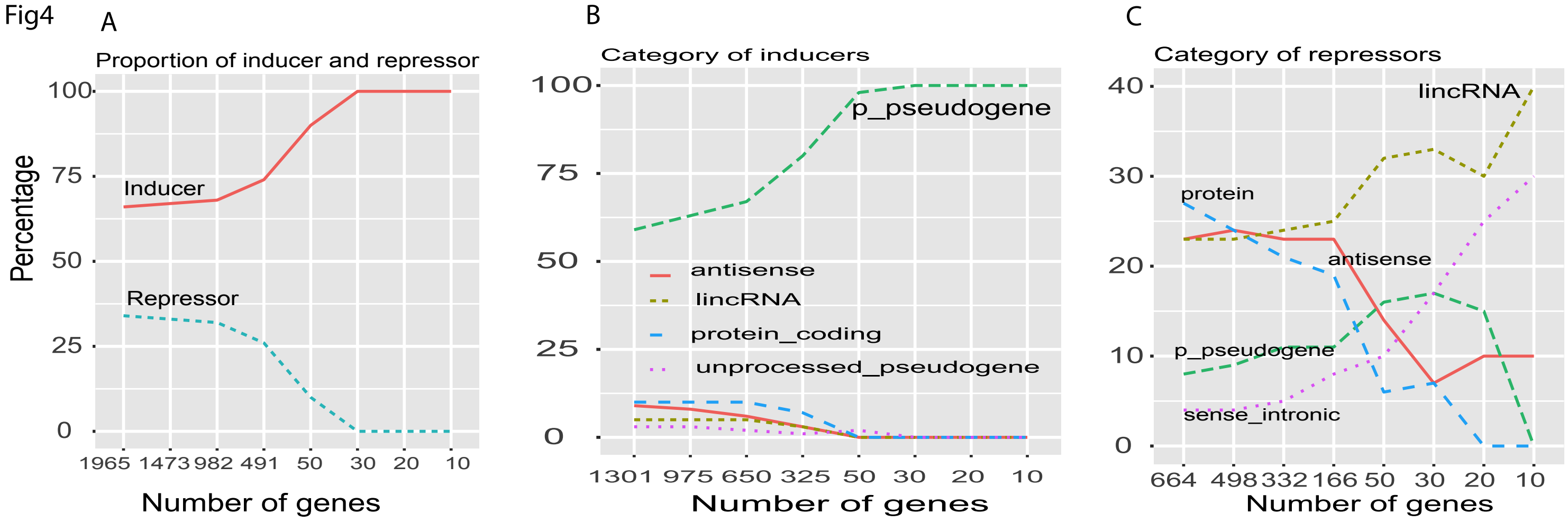

A
Proportion of inducer and repressor
Percentage
100
75
50
25
0
Inducer
Repressor
1965 1473 982 491 50 30 20 10
Number of genes
B
Category of inducers
100
75
50
25
0
p_pseudogene
antisense
lincRNA
protein_coding
unprocessed_pseudogene
1301 975 650 325 50 30 20 10
Number of genes
C
Category of repressors
40
30
20
10
0
lincRNA
protein
antisense
p_pseudogene
sense_intronic
664 498 332 166 50 30 20 10
Number of genes

Fig5

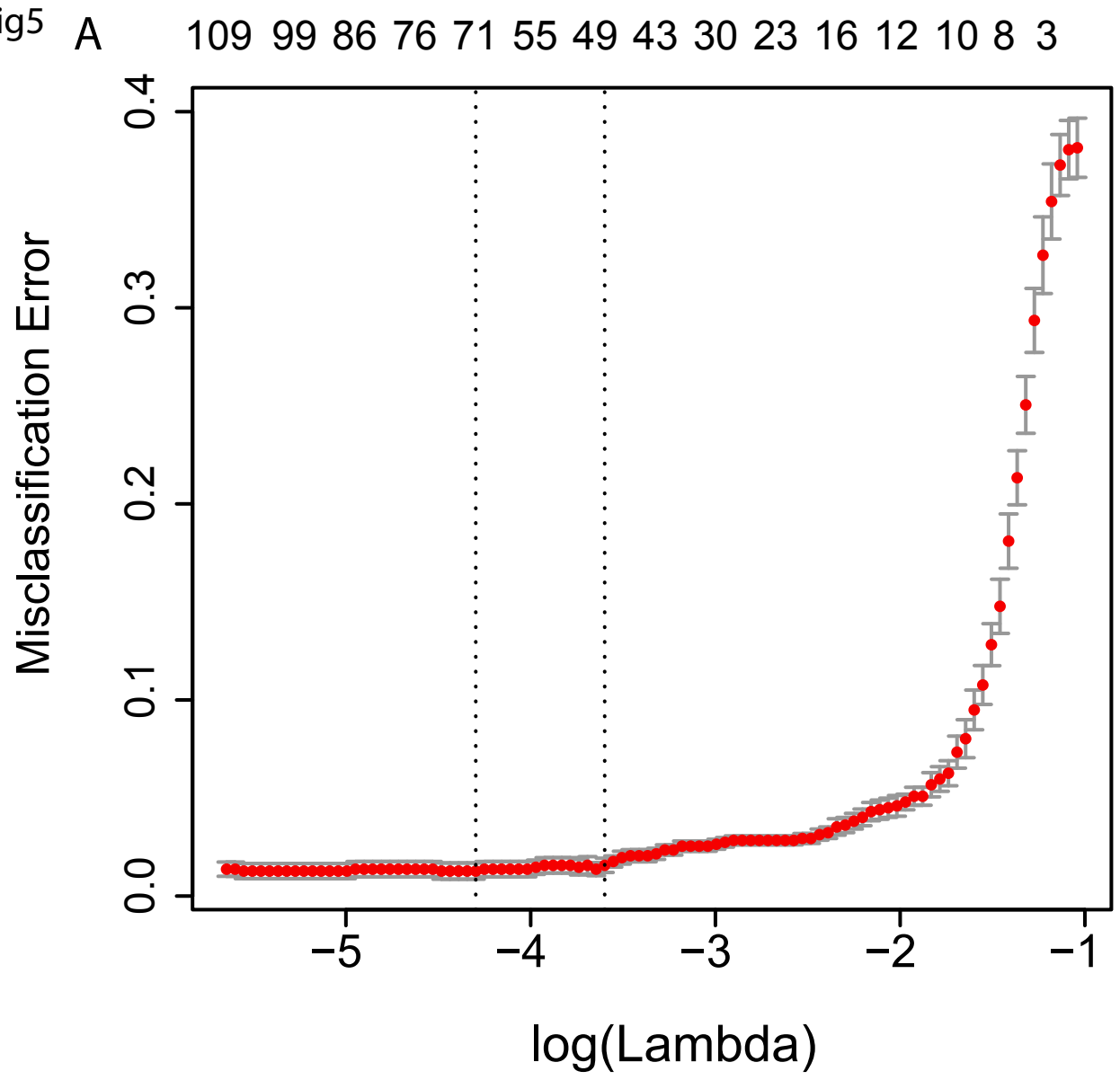

A
109 99 86 76 71 55 49 43 30 23 16 12 10 8 3
Misclassification Error
0.0
0.1
0.2
0.3
0.4
−5
−4
−3
−2
−1
log(Lambda)

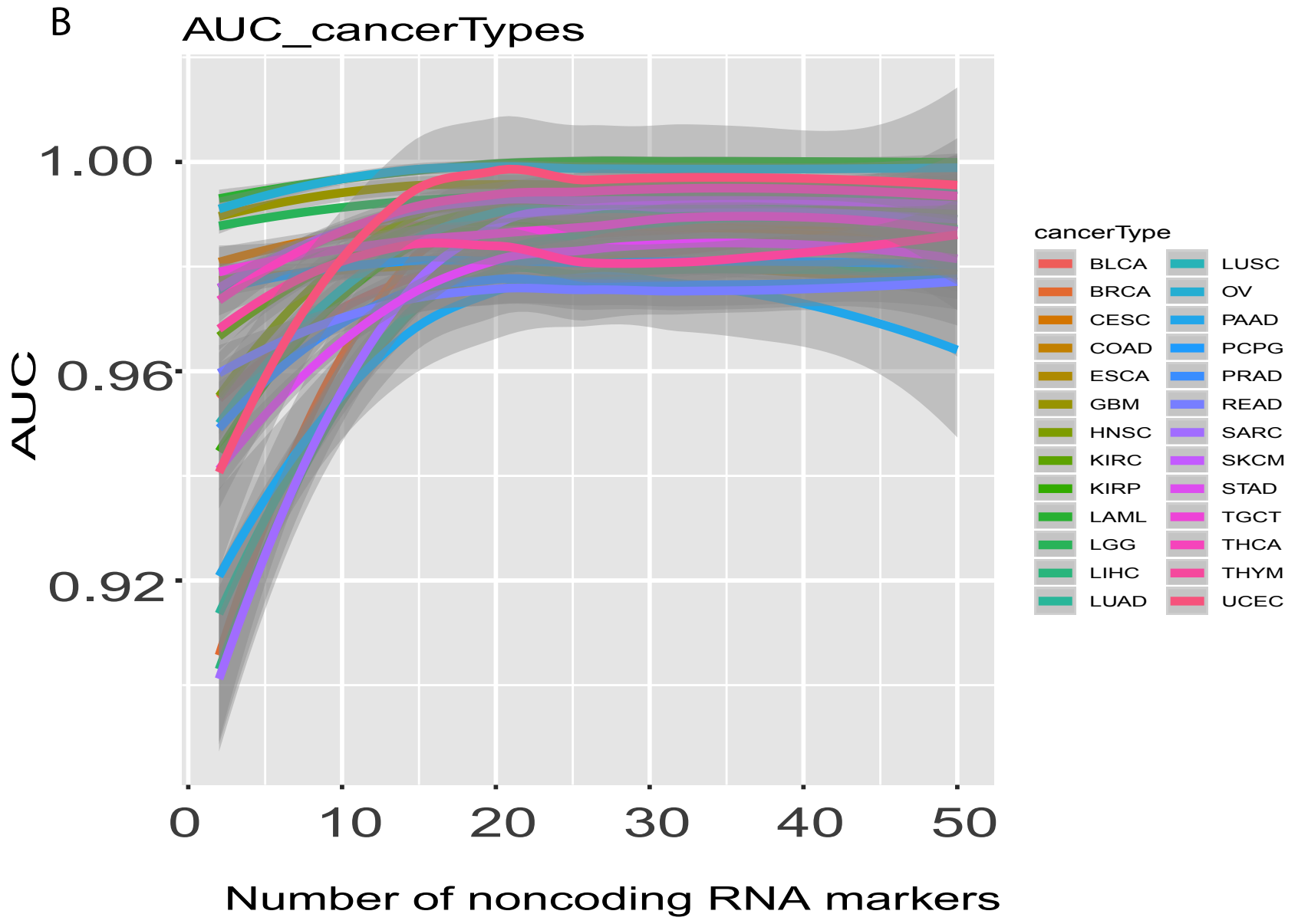